\documentclass[11pt]{article}
\usepackage{amsmath}
\usepackage{graphicx}
\usepackage{amssymb}
\usepackage{amsfonts}
\usepackage{latexsym}

\def\beq{\begin{eqnarray}}
\def\eeq{\end{eqnarray}}
\def\ln{\,\mbox{ln}\,}
\def\Ln{\,\mbox{Ln}\,}
\def\Det{\,\mbox{Det}\,}
\def\det{\,\mbox{det}\,}

\def\Tr{\,\mbox{Tr}\,}

\def\al{\alpha}
\def\be{\beta}

\def\ga{\gamma}

\def\ep{\epsilon}
\def\ze{\zeta}

\def\la{\lambda}
\def\na{\nabla}

\def\Ga{\Gamma}

\def\La{\Lambda}

\begin{document}

\hfill {\sl Preprint numbers: DF-UFJF/01-2009; \ \ 0904.4171 (hep-th)}
\vskip 4mm

\begin{center}
{\large \sc Exact formfactors in the one-loop curved-space QED
\\
and the nonlocal multiplicative anomaly}
\vskip 4mm

{\small \bf Bruno Gon\c{c}alves}
\footnote{brunoxgoncalves@yahoo.com.br}, 
\ 
{\small\bf Guilherme de Berredo-Peixoto}
\footnote{guilherme@fisica.ufjf.br}, 
\ 
{\small \bf Ilya L. Shapiro}
\footnote{Also at Tomsk State
Pedagogical University, Russia,  \ shapiro@fisica.ufjf.br}
\vskip 3mm

{\small\it Departamento de F\'{\i}sica, ICE, 
Universidade Federal de Juiz de Fora, 
\\
Juiz de Fora, CEP: 36036-330, MG,  Brazil}

\vskip 6mm

{\large\bf Abstract}
\end{center}

\vskip 2mm

\begin{quotation}
\noindent
The well-known formula $\,\det(A\cdot B)=\det A \cdot \det B\,$
can be easily proved for finite dimensional matrices but it may 
be incorrect for the functional determinants of differential 
operators, including the ones which are relevant for Quantum 
Field Theory applications. Considerable work has been done to 
prove that this equality can be violated, but in all previously 
known cases the difference could be reduced to renormalization 
ambiguity. We present the first example, where the difference 
between the two functional determinants is a {\it nonlocal} 
expression and therefore can not be explained by the 
renormalization ambiguity. Moreover, through the use of other 
even dimensions we explain the origin of this difference at 
qualitative level. 
\vskip 2mm

{\bf Pacs:} 
04.62.+v;    
11.15.Kc;   
11.10.Kk;	
12.20.Ds	


\end{quotation}
\vskip 8mm

\section{Introduction}

The one-loop calculations have a prominent role in Quantum 
Field Theory (QFT) and in many of its most relevant 
applications. In the background field method the one-loop 
contributions can be always reduced to the derivation of 
$\Ln \Det {\hat H}$ of the operator ${\hat H}$, which is 
typically a bilinear form of the classical action with 
respect to the quantum fields. The operator ${\hat H}$ 
usually depends on the background fields (which may be 
just external fields). As a result the operation of 
taking the functional determinant of such an operator 
is mathematically nontrivial due to the infinite dimension 
of the corresponding matrix representation. In particular, 
relations such as  
\beq
\Det({\hat A}\cdot {\hat B})&=&\Det{\hat A}\cdot\Det{\hat B}
\qquad \mbox{and}
\cr
\Ln \Det {\hat A} &=& \Tr \Ln  {\hat A}\,,
\label{multi}
\eeq
which are certainly valid for the finite dimensional matrices 
should be, in principle, proved or taken by faith in QFT. 
There is indeed another possibility that these relations
can be {\it disproved} and, according to mathematical logic, 
this can be done by means of at least one single nontrivial 
counterexample. For instance, that could mean a couple of 
operators, ${\hat A}$ and ${\hat B}$, for which the first 
relation in (\ref{multi}) would be violated. Such a situation 
was called {\it multiplicative anomaly} (MA) \cite{MA-0,MA-1}.

Considerable efforts have been applied to find an 
example where the first equality (\ref{multi}) would be
violated, but until now in all cases the difference was 
likely caused by the renormalization ambiguity only 
\cite{MA-2,MA-3,MA-4}. This means that when one imposes the 
renormalization conditions to the three operators ${\hat A}$, 
${\hat B}$ and ${\hat A}\cdot{\hat B}$, there may be a
difference due to the independence of these renormalization 
conditions for the distinct operators. In particular, such 
a situation can take place when the functional determinants 
are defined by means of the generalized $\ze$-function 
\cite{zeta}, because this approach ``hides'' the 
divergences and provides the regularized and renormalized 
result automatically. Then the $\mu$-dependence should be 
implemented artificially and this opens the way for the  
MA. The example of such a situation has been analyzed in 
detail in \cite{phase,MA-2}. If we consider, for example, 
the 
$\Ln\Det \big(\Box+M_1^2\big) \cdot \big(\Box+M_2^2\big)$ 
on de Sitter background, the result will 
be a functional which depends on some constant parameters, 
namely on $M_{1,2}^2$ and on the scalar curvature $\La$. 
Furthermore, 
this expression has dimension four. As a result it has 
exactly the same structure as the counterterms and, 
therefore, it is a subject of the renormalization 
ambiguity. Thus, it is very difficult to make 
positive conclusion on the existence of the MA based
on such calculations. In order to establish the existence 
of the MA one needs to find it in such a finite sector of 
the effective action which can be clearly different from 
the counterterms. 

The purpose of the present letter is to present an example 
of another sort, that means the {\it nonlocal} MA  which 
is not reduced to the renormalization ambiguity. In order 
to construct such example we consider one of the most 
familiar theories, that is the usual spinor QED. We 
consider a curved space-time, but the effect can be 
observed even in flat space-time. This letter represents 
a short communication devoted to the MA and we leave 
technical details to the parallel publication \cite{Large},
devoted to the general investigation of quantum violation 
of conformal invariance for electromagnetic fields. 

\section{Photon formfactors in the 1-loop QED}

Consider the problem of deriving the correction to the 
electromagnetic field propagation from the single loop 
of a Dirac fermion. The Euclidean action has the form 
\beq
 S = \int d^4x\sqrt{g}\Big\{\bar{\psi}
 \big(i\ga^\mu\na_\mu + e \ga^\mu A_\mu + M\big)\psi
 - \frac{1}{4}\,F^{\mu\nu}F_{\mu\nu}\Big\}.
\nonumber
\eeq
The one-loop effective action (EA) in the metric and 
electromagnetic sectors can be defined via the path 
integral 
\beq
e^{i \Gamma[g_{\mu\nu},\,A_\mu]}=
\int{D\psi D\bar{\psi}\, e^{i S}}\,.
\label{eaction}
\eeq
In the conventional form we find (see, e.g., \cite{book})
\beq
{\bar \Ga}^{(1)} &=& \,-\,\frac{1}{2}\,\Ln \Det\,\hat{H}\,,
\cr
\hat{H} &=& i\,\big(\ga^\mu\na_\mu -ie \ga^\mu A_\mu-iM\big)\,.
\label{1-loop}
\eeq
In order to use the heat kernel method,
one has to multiply $\hat{H}$ by a conjugate 
operator $\hat{H}^*$, such that the product has the 
form 
$\hat{H}\hat{H}^*={\widehat{\Box}}+2{\hat h}^\mu\na_\mu
+ {\hat \Pi}$. The point is that the choice of the 
conjugate operator $\hat{H}^*$ is not unique. Here we
consider the two following choices: 
\beq
\hat{H}^{*}_1 &=& 
-i\gamma^\mu \na_\mu + M-e\gamma^\nu A_\nu
\qquad
\mbox{and}
\cr
\hat{H}^{*}_2 &=& -i\gamma^\mu \na_\mu + M\,.
\label{H2}
\eeq
In what follows the difference between the 
$\hat{H}^{*}_1$ and $\hat{H}^{*}_2$ cases will be named 
``scheme dependence''. The question is whether the 
$F^{\mu\nu}(...)F_{\mu\nu}$-type terms calculated 
via the expressions
$\Ln \Det\,\hat{H}\hat{H}^{*}_1$ and 
$\Ln \Det\,\hat{H}\hat{H}^{*}_2$ are the same or not. 
In both cases we assume 
\beq
\Ln\Det \hat{H} = \Ln\Det \big(\hat{H}\hat{H}^{*}\big) 
- \Ln\Det \hat{H}^{*}\,. 
\label{hipo}
\eeq
Now, in the first case the contributions of 
$\Ln\Det \hat{H}$ and $\Ln\Det \hat{H}_1^{*}$ are
equal \cite{Guilherme}, so in fact we can take 
$$
\Ln\Det \hat{H} 
= \frac{1}{2}\,\Ln\Det \big(\hat{H}\hat{H}_1^{*}\big)\,.
$$ 
In the second case the expression $\Ln\Det \hat{H}_2^{*}$ 
does not depend on $A_\mu$ and therefore the 
$F^{\mu\nu}(...)F_{\mu\nu}$-type terms satisfy 
(using obvious notations) the relation 
$$
\Ln\Det \hat{H}\Big|_{FF} 
= \Ln\Det \big(\hat{H}\hat{H}_2^{*}\big)\Big|_{FF}\,.
$$
So, if the first identity from (\ref{multi}) holds, 
we are going to meet the two equal expressions, 
\beq
\frac{1}{2}\,\Ln\Det \big(\hat{H}\hat{H}_1^{*}\big)\Big|_{FF} 
\,=\,\Ln\Det \big(\hat{H}\hat{H}_2^{*}\big)\Big|_{FF}\,,
\label{criter}
\eeq
but if (\ref{criter}) does not hold, (\ref{multi}) is 
violated. We will show that in fact the two expressions 
have different finite parts despite the divergent parts 
being equal. Moreover in the case of $\hat{H}^{*}_2$ the 
gauge invariance is violated in the finite part of EA. 
Let us note that the last occurrence can be seen as one 
more confirmation of the MA. The reason is
that the expression $\Ln \Det\,\hat{H}$ is gauge invariant 
by construction (we assume invariant regularization)
and the expression $\Ln \Det\,\hat{H}^{*}_2$ does not 
depend on the gauge field $A_\mu$ and hence it is also 
gauge invariant. Hence, if $\Ln \Det\,\hat{H}\hat{H}^{*}_2$
is non-invariant, then 
\\ \\
\centerline{
$\Ln \Det\,\hat{H}\hat{H}^{*}_2 \neq
\Ln \Det\,\hat{H} + \Ln \Det\,\hat{H}^{*}_2$
}
\\ \\
and we meet one more evidence of the MA. 

Let us see whether the situation described above 
really takes place. In order to calculate 
$\Ln \Det\,\hat{H}\hat{H}^{*}_1$ and 
$\Ln \Det\,\hat{H}\hat{H}^{*}_2$ we use the heat 
kernel solution \cite{bavi90} which was earlier applied 
to the derivation of formfactors in the gravitational sector
\cite{apco,fervi}. 
Let us note also that the same result can be achieved via 
the Feynman diagrams \cite{apco}.

The one-loop quantum correction for the $\hat{H}^{*}_1$ 
case has the form
\beq
{\bar \Ga}^{(1)}\Big|_{FF}
&=& - \frac{e^2}{2(4\pi)^2}\int d^4x \sqrt{g}
F_{\mu\nu} 
\Big[\frac{2}{3\ep}+k^{FF}_1(a)\Big] F^{\mu\nu},
\nonumber
\\ 
\nonumber 
\\
\mbox{with} &\,& k^{FF}_1(a) =
Y\Big(2-\frac{8}{3a^2}\Big)-\frac{2}{9}\,,
\label{FF1}
\\ 
\nonumber 
\\
\nonumber 
\mbox{where} &\mbox{we}& \mbox{used the following notations:}
\nonumber 
\\ 
\nonumber 
\\
\nonumber 
Y &=& 1-\frac{1}{a}\ln\,\Big(\frac{2+a}{2-a}\Big)\,,
\qquad a^2 = \frac{4\Box}{\Box - 4m^2}\,. 
\nonumber 
\nonumber
\eeq
For the $\hat{H}^{*}_2$ case we meet a different result, 
namely (we do not use the notation $\big|_{FF}$ here 
because there are other ${\cal O}(A^2)$-terms)
\beq
{\bar \Ga}^{(1)}\Big|_{AA}
&=& 
- \frac{e^2}{2(4\pi)^2} \int d^4x \sqrt{g}\,\Big\{
F_{\mu\nu} \,\Big[\frac{2}{3\,\ep}+k^{FF}_2(a)\Big] F^{\mu\nu} 
\\
\label{FF2}
&+& 
\na_\mu A^\mu 
\Big[Y \big( \frac{8}{3a^2} - 2\big) 
+ \frac {2}{9} \Big] \na_\nu A^\nu 
\\
\nonumber
&+& R_{\mu\nu} \,\Big[\frac{8Y}{3a^2} 
+ \frac{2}{9} \Big] A^\nu A^\mu 
\,+\, A^\nu A^\mu 
\Big[\frac{8Y}{3a^2} + \frac{2}{9}\Big] 
R_{\mu\nu} 
\\ \nonumber &+& 
\na_\mu A^\nu \Big[ \frac{16Y}{3a^2} 
+ \frac {4}{9} \Big] \na_\nu A^\mu 
+ {\cal O}(R\cdot A\cdot A) \Big\}\,,
\\
\nonumber
&\mbox{where}& \,\,\,
k^{FF}_2(a) =
Y\Big(1+\frac{4}{3a^2}\Big)+\frac{1}{9}\,,
\eeq
and ${\cal O}(R\cdot A\cdot A)$ are terms proportional 
to scalar curvature. 

In the expressions (\ref{FF1}) and (\ref{FF2}),
$\,\ep\,$ is the parameter of dimensional regularization
$$
\frac{1}{\ep}=\frac{2}{4-d}
+\ln \Big(\frac{4\pi \mu^2}{m^2}\Big) - \ga
\,,
\qquad \ga=0.5772\,...\,
\,.
$$
It is easy to see that the divergences are exactly the 
same in the two expressions but, at the same time, 
the finite parts indicate the presence of MA. In fact,  
the situation is exactly as it was described above. 
In the divergent parts of the two formulas (\ref{FF1}) 
and (\ref{FF2}) there is no scheme dependence, 
while the finite nonlocal parts of these expressions
do differ and, also, (\ref{FF2}) is not gauge invariant. 
The scheme dependence can not be eliminated by adjusting 
the renormalization condition, because the last does 
not concern the nonlocal part of EA. So, we have 
confirmed the existence of MA for the Dirac operator. 
However, in this situation the interested 
reader has the right to ask natural questions like: 
``Is it all correct?'' and ``Why does the MA take place?'' 
Of course, the first question can be addressed only 
through a clear answer to the second one, and we 
will present such an answer in the next section. 

\section{${\hat a}_n$ coefficients and the origin of MA}

In order to understand the origin of the MA, let us 
remember that the heat kernel solution of \cite{bavi90} 
is a sum of the series of the coincidence limits of the 
Schwinger-DeWitt coefficients $a_n(x,x^\prime)$. The 
equal divergences of the two effective actions (\ref{FF1}) 
and (\ref{FF2}) mean that the coefficients $a_2$ of 
the two operators do coincide in the four-dimensional 
space. The distinct finite parts 
mean that some other coefficients are in fact different.
Therefore the natural way to check the correctness of 
the results (\ref{FF1}) and (\ref{FF2}) is to calculate 
the coincidence limit of some other coefficient, e.g., 
$a_1(x,x^\prime)$, or $a_3(x,x^\prime)$. Before we begin
our calculations, let us imagine what should we expect as
a possible output. For this end it is most interesting to 
consider an arbitrary dimension $d$ of space-time. The 
$4d$ case considered above has shown that the divergent 
part of the effective action is scheme-independent and 
thus universal. Mathematically, there is nothing special 
about $4d$, so we can expect that this universality holds 
also in other even dimensions. 

Let us note that the expression 
${\hat a}_k=\Tr \lim\limits_{x^\prime\to x}a_k(x,x^\prime)$ 
with $k=1$ corresponds to the UV divergence of EA in $2d$,
with $k=2$ in $4d$, with $k=3$ in $6d$ etc. Therefore 
the universality of the UV divergences implies that 
${\hat a}_1$ is universal in $2d$, ${\hat a}_2$ in $4d$, 
${\hat a}_3$ in $6d$ etc. The most interesting 
moment in this story is that the universality of the 
Schwinger-DeWitt coefficients in the ``right'' 
dimensions automatically implies the non-universality 
of the overall finite contributions in {\it any} 
particular dimension! The point is that the general
expression for the coincidence limit 
$\lim\limits_{x^\prime\to x}a_k(x,x^\prime)$ does not 
depend on $d$, but the corresponding functional trace
${\hat a}_k$ does. As a result, if the two traces are
equal in the ``right'' dimensions, they are unlikely to 
be equal in other dimensions. 
For instance, all terms except ${\hat a}_2$ are
scheme-dependent in $4d$, and therefore the sum of the 
series made out of these terms is also not universal. 
Indeed, this is exactly what we observe in the 
formfactors (\ref{FF1}) and (\ref{FF2}) calculated 
within the two distinct schemes.  

Let us verify that the considerations presented above 
are correct. We start from the evaluation of ${\hat a}_1$
in $2d$. We know that the $\,{\hat a}_1=\int\sqrt{g}\hat{P}$,
where $\hat{P}$'s in the two cases are given by the expressions 
\beq
\hat{P}_1 &=& - \frac{1}{12}\,R + M^2 
- \frac{ie}{2}\gamma^\mu \gamma^\nu F_{\mu \nu}\,,
\nonumber 
\\
\hat{P}_2 &=& - \frac{1}{12}\,R
- \frac{ie}{4}\gamma^\mu \gamma^\nu F_{\mu \nu}
+ M^2 + eM \gamma^\mu A_\mu  \cr
\nonumber 
\\
&+&
\frac{ie}{2}\, (\na^\mu A_\mu)-
\frac{(d-2)}{4}e^2 A^\nu A_\nu\,.
\label{PS}
\eeq
It is easy to see that the difference 
between the two traces is reduced to the total derivative 
in $2d$, while in other dimensions it is more significant. 
Furthermore, only in $2d$ the $\Tr \hat{P_2}$ is a
gauge invariant expression. Let us note that the difference 
in total derivative may indicate some real thing for 
the finite part of EA, but not for renormalization. 
Therefore the general expectation described above is 
completely confirmed in the ${\hat a}_1$ case. 
We leave it as an exercise to the reader to check that 
the situation is the same for the ${\hat a}_2$ coefficients, 
where the two schemes give equal results in the $4d$ case 
and distinct results for $d\neq 4$ cases. 

As a last test, let us now consider the ${\hat a}_3$ 
coefficient. 
Within the first calculational scheme with $\hat{H^{*}_1}$
of (\ref{H2}), we just confirm the known result of Ref. 
\cite{Drummond},
\beq
{\hat a}_3^{(1)}\Big|_{AA} 
&=& \frac {d\,e^2}{360} 
\,\big(\,2\,R_{\mu\nu\al\be} \, F^{\mu\nu}F^{\al\be} 
\,-\, 26\,R^\al_\nu \,F^{\mu\nu} \,F_{\mu\al} 
\cr
&+& 24\,\na_\nu F^{\mu\nu}\, \na_\al F_\mu^{\,\,\,\,\al}
\,+\, 5\,R \, F^{\mu\nu} \, F_{\mu\nu}\,\big)\,.
\label{a3-1}
\eeq
The expression ${\hat a}_3^{(2)}$ for the second scheme, 
with $\hat{H^{*}_2}$, is rather bulky
\big[here $(\na A)=(\na_\mu A^\mu)$\big]:  
\beq
{\hat a}_3^{(2)}\Big|_{AA} 
&=& 
\frac{de^2}{2880} \Big\{ 
120(\na A)\Box(\na A)
-60 F_{\mu\nu}\Box F^{\mu\nu}
\cr
\cr
&-&
24\na_\nu F^{\mu\nu}\na^\al F_{\mu\al}
+ 24(\Box A^\al)\big[(d-3)(\Box A_\al) 
+ 2\na_\al(\na A)\big]
\cr
\cr
&-&
24(\na_\al\na_\mu A_\be)\left[(\na^\be\na^\mu A^\al) 
- (\na^\al\na^\mu A^\be)\right] 
\cr
\cr
&+&
A^\mu A_\mu 
\left[(18-7d)R^2_{\mu\nu\al\be}
-8(9-d)R^2_{\mu\nu}-6(5-d)R^2\right]
\cr
\cr 
&+& 8R_{\mu\nu\al\be}\big[
4 (\na^\al A^\nu)(\na^\mu A^\be)
-8 F^{\mu\nu}F^{\al\be}
- 3(d-4) (\na^\mu A^\al)(\na^\nu A^\be)
\cr
\cr
&-&
 R^{\la\nu\al\be}A^\mu A_\la
+ 10 R^{\mu\be}  A^\al A^\nu
\big]
\cr
\cr
&+&  16 R_{\mu\nu} 
\big[
10 (\na A)(\na^\mu A^\nu)
\,+\,
(\na^\al A^\mu)(5\na_\al A^\nu
- 2\na^\nu A_\al)
\cr
\cr
&-&
(d-5)(\na^\mu A^\al)(\na^\nu A_\al)
-2 R^\mu_{\,\,\,\al}\,A^\al A^\nu
\big]
\cr
\cr
&+&  10R \big[
2(d-5)(\na_\mu A_\nu)(\na^\mu A^\nu)
- 2 (\na A)^2
+ 3 F_{\mu\nu} F^{\mu\nu}
+ 2 R_{\mu\nu} A^{\mu}A^{\nu}
\big]
\cr
\cr
&-&
 12(d-2)A^\al A_\al \Box R 
\cr
\cr
&-& 48(\na^\al R_{\mu\nu\al\be})
(\na^\nu A^\be A^\mu)
- 24(\na^\nu R)
[(\na_\nu A^\al A_\al) - (\na_\al A^\al A_\nu)]\Big\}\,.
\label{g2renAAAAA}
\eeq
It is easy to check that, in $4d$, 
the formulas (\ref{a3-1}) and (\ref{g2renAAAAA}) do
coincide with the third orders of the expansions of the 
complete expressions (\ref{FF1}) and (\ref{FF2}), 
correspondingly. This correspondence serves as an 
independent verification for the correctness of our
formfactors (\ref{FF1}) and (\ref{FF2}). 

The comparison of the expressions (\ref{a3-1}) 
and (\ref{g2renAAAAA})
shows that, in the flat space limit, the ${\hat a}_3^{(2)}$ 
does coincide with ${\hat a}_3^{(1)}$ in $6d$ and 
only in $6d$. Furthermore, we could prove that the terms 
porportional to $R\,F^{\mu\nu}F_{\mu\nu}$ in two 
expressions ${\hat a}_3^{(2)}$ and ${\hat a}_3^{(1)}$ 
coincide (up to total derivatives) on dS/AdS background. 
In any other dimension the gauge invariance is broken even 
in the flat space background, as it was expected from 
general arguments given above. The difference between 
${\hat a}_3^{(2)}$ and ${\hat a}_3^{(1)}$ is precisely 
the one which can be observed between the first terms of 
expansion of the general expressions (\ref{FF1}) and
(\ref{FF2}). At that point we can say that our general 
arguments concerning the origin of the MA is very well
supported by direct calculations of the first three
Schwinger-DeWitt coefficients.   

\section{Appelquist and Carazzone theorem}

Let us look 
at the UV and IR limits of the physical $\be$-functions for the 
charge $e$. Starting from the expressions (\ref{FF1}) and 
(\ref{FF2}), correspondingly, we arrive at the following 
expressions for the $\be$-functions: 
\beq
\be_e^1= \frac{e^3\left[ 48  - 20 a^2 
+ 3(a^2-4)^2 (1-Y)\right]}{6a^2(4\pi)^2}\,, 
\label{beta1}
\eeq
versus
\beq
\be_e^2= \frac{e^3
\left[ 3(a^4-16) - 4a^2 (12+a^2)(1-Y) \right]}{12a^3(4\pi)^2}\,.
\label{beta2}
\eeq

In the high energy limit, when $p^2 \gg m^2$, 
$a \rightarrow 2$ and the two expressions give identical 
results, which also coincides with the one from the minimal 
subtraction scheme (up to a small correction), 
\beq
\be_e^{UV} \,=\,\frac{4\,e^3}{3\,(4\pi)^2}\,
+ \,{\cal O}\Big(\frac{m^2}{p^2}\Big)\,,
\label{beta1-UV}
\eeq
However, at the low-energy end the results are different, 
namely
\beq
\be_e^{1\,\,IR} \,=\, \frac{e^3}{(4\pi)^2}\,\cdot\,
\,\frac{4\,M^2}{15\,m^2} \,\,
+ \,\,{\cal O}\Big(\frac{M^4}{m^4}\Big)
\label{beta1-IR}
\eeq
for the first scheme $\hat{H}^*_1$, and 
\beq
\be_e^{2\,\,IR} \,=\, \frac{e^3}{(4\pi)^2}\,\cdot\,
\frac{1}{5} \,\,
\frac{M^2}{m^2} \,\,
+ \,\,{\cal O}\Big(\frac{M^4}{m^4}\Big)
\label{beta2-IR}
\eeq
for the second one, with $\hat{H}^*_2$. Thus we met a 
scheme ambiguity, also, in the decoupling theorem \cite{AC}. 

In order to better understand the sense of the MA and 
the above difference in the $\be$-functions, we can look 
at the lowest order term in the EA, where the difference
shows up, 
\beq
\frac{1}{30} \cdot \int d^4x \sqrt{g}\,
F^{\mu\nu} \Big(\frac{\Box}{m^2}\Big) F_{\mu\nu}
\label{FF-EA}
\eeq  
In the flat space-time, one can easily use integrations
by parts to show that this term is proportional to the
Maxwell equations, $(\na_\mu F^{\mu\nu})^2$. Hence this
term will not influence the equations of motion in 
flat space in the ${\cal O} (e^2)$ approximation 
\cite{Drummond}. However, the situation gets changed when 
we deal with the curved space. In this case we meet a 
difference thar is proportional to curvatures, due to 
the relation
\beq
F^{\mu\nu}\, \Box \,F^{\mu\nu} &=& 
-2 \na_\nu F^{\mu\nu} \na_\la F_\mu^{\,\,\,\la} \,+\,
2 R_{\la\nu} F^{\mu\nu} F_\mu^{\,\,\,\la} 
\cr
&-& R_{\al\be\mu\nu} F^{\mu\nu} F^{\al\be}\,.
\eeq
It is important that this difference is also confirmed 
by the derivation of the ${\hat a}_3$ coefficient
described in the previous section.

\section{Conclusions}

We have calculated the formfactor in the electromagnetic 
sector of QED in curved space-time and found that this
quantum correction depends on the choice of the 
calculational scheme (\ref{H2}). Thus we have proven 
the existence of the nonlocal and renormalization 
independent MA in quantum field theory. One of the 
consequences of this anomaly is the ambiguity in the 
prediction of the decoupling 
theorem 
\cite{AC}, which provides two different coefficients 
of the quadratic decoupling law at low energies. 

The MA in the electromagnetic formfactor means that the 
off-shell EA possesses some new important ambiguity. 
One can use the Maxwell equation and show that in the 
flat space the ambiguous terms disappear on shell. 
However, this does not happen in curved space where 
we meet a real ambiguity proportional to the 
$RFF$-terms. 

How should we interpret the existence of MA? In fact, 
the EA is always ambiguous to some extent. For instance, 
there is a strong dependence on the choice of
parametrization for the quantum field \cite{tyutin} 
which becomes relevant beyond the leading-log
approximation. Perhaps, from the practical viewpoint 
the best option is to follow the most natural approach 
and, for instance, take the most natural parametrization 
of quantum fields and the most natural and symmetry 
preserving scheme of calculation. On the other hand, 
it is always good to be aware on the real features of
the utilized formalism, and from this perspective it 
is indeed important to know that the MA is a real thing.   
\vskip 3mm

\section*{Acknowledgments}
Authors are grateful to CNPq, FAPEMIG, FAPES and ICTP 
(I.Sh.) for support.


\end{document}